# Optimized time-lapse acquisition design via spectral gap ratio minimization

Yijun Zhang[1], Ziyi Yin[1], Oscar Lopez[2], Ali Siahkoohi[3],
Mathias Louboutin[1], Rajiv Kumar[4], Felix J. Herrmann[1]
[1] Georgia Institute of Technology, [2] Harbor Branch Oceanographic Institute,
[3] Rice University, [4] Schlumberger

**Abstract**

Modern-day reservoir management and monitoring of geological carbon storage increasingly call for costly time-lapse seismic data collection. In this letter, we show how techniques from graph theory can be used to optimize acquisition geometries for low-cost sparse 4D seismic. Based on midpoint-offset domain connectivity arguments, the proposed algorithm automatically produces sparse non-replicated time-lapse acquisition geometries that favor wavefield recovery.

# Introduction

Time-lapse seismic data acquisition is a costly but crucial endeavor for reservoir management and monitoring of geological carbon storage. While sparse randomized collection of seismic data can lead to major improvements in acquisition productivity [Herrmann and Hennenfent, 2008, Hennenfent and Herrmann, 2008, Herrmann, 2010, Mosher et al., 2014], systematic approaches to performance prediction, other than computationally expensive simulation-based studies, are mostly lacking. Besides, acquisition optimization approaches, such as minimizing the mutual coherence [Tang et al., 2008, Mosher et al., 2014, Obermeier and Martinez-Lorenzo, 2017] or minimizing the spectral gap ratio [SGR, López et al., 2022, Zhang et al., 2022], do not handle the unique challenges of time-lapse seismic data acquisition.

To meet these challenges, inversion with the joint recovery model [JRM, Oghenekohwo et al., 2017, Wason et al., 2017] will be combined with automatic binary sampling mask generation driven by SGR minimization [Zhang et al., 2022]. We opt for the JRM because it inverts baseline and monitor surveys jointly for the common component, which contains information shared between the surveys, and innovations with respect to the common component. Since the fictitious common component is observed by all surveys, its recovery improves when the time-lapse surveys contain complementary information. This is the case when sparse surveys are not replicated [Oghenekohwo et al., 2017, Wason et al., 2017] or when the time-lapse datasets contain independent noise terms [Tian et al., 2018]. In either case, the JRM leads without insisting on replication of the surveys to high degrees of time-lapse repeatability both in the data [Oghenekohwo et al., 2017, Wason et al., 2017] and image space [Yin et al., 2023]. It also yields better interpretability of time-lapse field data [Wei et al., 2018].

As demonstrated by this letter, including the common component offers additional advantages when optimizing time-lapse acquisition via SGR minimization. To demonstrate this, we first explain the relationship between the SGR and connectivity within graphs associated with binary sampling masks. Next, we describe how this connectivity, which favors wavefield reconstruction, can be improved by minimizing the SGR via optimization. To enhance inversion of time-lapse data with the JRM, a new optimization objective will be introduced that contains SGRs of the common component and of the baseline/monitor surveys. After a brief discussion on minimizing this objective with simulated annealing, the proposed methodology for automatic time-lapse binary mask generation is numerically validated on realistic synthetic 2D data.

## Optimized time-lapse acquisition

While the SGR has been used successfully to predict and improve the performance of wavefield reconstruction, it has not yet been used to optimize time-lapse acquisition. After briefly discussing the SGR and JRM, we introduce our methodology to optimize time-lapse data acquisition.

### The spectral gap ratio

As shown by López et al. [2022], the success of seismic wavefield reconstruction via universal matrix completion [Bhojanapalli and Jain, 2014] can be predicted by the ratio of the first two singular values of binary sampling masks, $\sigma_2(\mathbf{M})/\sigma_1(\mathbf{M}) \in [0, 1]$ where $\mathbf{M}$ is a binary matrix with 1's where data is sampled and with 0's otherwise. This ratio is known as the spectral gap ratio (SGR) and provides a cheap-to-compute quantitative measure to predict recovery quality. The smaller the SGR, the better the connectivity within graphs spanned by binary sampling masks. Improved connectivity leads to improved wavefield recovery [López et al., 2022]. While useful, the SGR itself is not constructive because it does not produce sampling masks with small SGRs. With simulated annealing, Zhang et al. [2022] came up with a practical algorithm to generate acquisition geometries with small SGRs. In this work, we extend this approach by optimizing sparse geometries for time-lapse data acquisition.

### Optimized sampling mask generation

Given an initial binary mask, $\mathbf{M} \in \{0,1\}^{n_s \times n_r}$, with $n_s$ sources and $n_r$ receivers, Zhang et al. [2022] proposed a methodology to minimize the SGR via

$$\underset{\mathbf{M}}{\text{minimize}} \quad \mathcal{L}(\mathbf{M}) \quad \text{subject to} \quad \mathbf{M} \in \mathcal{C}, \tag{1}$$

with the objective, $\mathcal{L}(\mathbf{M}) = \sigma_2(\mathbf{M})/\sigma_1(\mathbf{M})$, given by the SGR. To ensure feasibility of the optimized binary masks, the constraint, $\mathcal{C} = \bigcap_{i=1}^{3} \mathcal{C}_i$, is included, which consists of the intersection of the cardinality constraint, $\mathcal{C}_1 = \{\mathbf{M} \mid \#(\mathbf{M}) = \lfloor n_s \times \rho \rfloor \times n_r\}$, the binary mask constraint, $\mathcal{C}_2 = \{\mathbf{M} \mid \mathbf{M} \in \{0,1\}^{n_s \times n_r}\}$, and a constraint on the maximum gap size between consecutive samples, $\mathcal{C}_3 = \{\mathbf{M} \mid \text{maxgap}(\mathbf{M}) \leq \Delta\}$, where $\Delta$ is the maximal permitted gap size. By solving Equation 1, Zhang et al. [2022] produced binary sampling masks that improved wavefield reconstruction compared to masks generated with randomized jittered sampling [Hennenfent and Herrmann, 2008]. Figure 1 contrasts jittered with optimized sampling in the midpoint-offset domain, reducing the SGR from 0.333 to 0.196. The optimized mask increases the sampling at the near offsets where there are more ways to connect to midpoints, which favors wavefield reconstruction [López et al., 2022].

### Joint recovery model

Lowering costs while ensuring time-lapse repeatability are the main challenges in the design of seismic monitoring systems employed to optimize reservoir management and to safeguard geological carbon storage. Both challenges can be met by inverting sparsely sampled baseline and monitor data jointly. For time-lapse acquisition with a single monitor survey, this entails inverting

$$\mathbf{b} = \mathcal{A}(\mathbf{Z}) \quad \text{with} \quad \mathcal{A}(\cdot) = \begin{bmatrix} \mathcal{A}_1 & \mathcal{A}_1 & 0 \\ \mathcal{A}_2 & 0 & \mathcal{A}_2 \end{bmatrix} (\cdot). \tag{2}$$

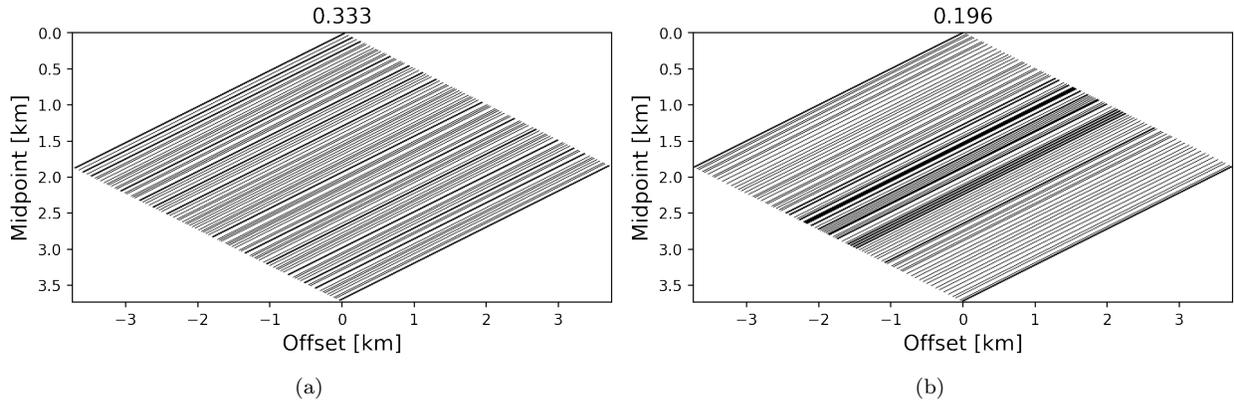

Figure 1: (a) Jittered versus (b) optimized sampling mask in the midpoint-offset domain.

In this JRM, the linear operators, $\mathcal{A}_j$, $j = 1, 2$, stand for the combined action of converting monochromatic time-lapse data from the midpoint-offset to the source-receiver domain, followed by trace collection with the acquisition geometries defined by the binary sampling masks, $\mathbf{M}_j$, $j = 1, 2$ with $j = 1$ and $j = 2$ masks for the baseline/monitor surveys. With this model, time-lapse data, $\mathbf{b}$, which contains the baseline, $\mathbf{b}_1$ and monitor data, $\mathbf{b}_2$, are linearly related to $\mathbf{Z}$, which contains matrices for the unknown densely sampled common component, $\mathbf{Z}_0$, and innovations with respect to this common component, $\mathbf{Z}_j$, $j = 1, 2$. Compared to recovering the baseline/monitor surveys separately, the JRM produces repeatable results from non-replicated [Oghenekohwo et al., 2017, Wason et al., 2017, Kumar et al., 2017], non-calibrated [Oghenekohwo and Herrmann, 2017], and noisy [Tian et al., 2018], time-lapse data. These enhanced results are due to the improved recovery of the fictitious common component.

**Time-lapse optimized mask generation**

Based on the success of the JRM, we carry the argument of minimizing the SGR a step further by optimizing this quantity for the baseline/monitoring surveys. Because $\mathbf{Z}_0$ is observed by both surveys, the set of sampling points, $\{\mathbf{M}_0\}$, equals the union $\{\mathbf{M}_0\} = \{\mathbf{M}_1\} \cup \{\mathbf{M}_2\}$. When surveys are replicated, $\{\mathbf{M}_0\} = \{\mathbf{M}_1\} = \{\mathbf{M}_2\}$. However, $\mathbf{M}_0$ becomes larger when the baseline and monitor surveys are not replicated explaining why the common component is better resolved when the surveys are not replicated.

While Equation 1 leads to improved sampling masks for individual surveys, it does not exploit the fact that the common component is observed by all surveys. For this reason, we propose an optimization procedure with respect to $\mathbf{M}_1$ and $\mathbf{M}_2$ with an objective that also includes the SGR for the common component. To avoid generation of poor sampling masks, we follow a mini-max principle where the maximum—i.e., the $\ell_\infty$-norm—of the SGRs for the common and innovation components is minimized. To compensate for likely smaller SGRs for the common component when the surveys do not overlap ($\#\{\mathbf{M}_0\} > \#\{\mathbf{M}_1\}, \#\{\mathbf{M}_2\}$), we also introduce a scaling. We base this scaling on the property [see Definition 3.1 in Bhojanapalli and Jain, 2014, Hoory et al., 2006] that the second singular value of $d$-regular graphs—i.e., seismic sampling masks with $d$ non-zero entries per midpoint or offset— scales with $\sqrt{d}$. Given this scaling, we propose to minimize the following constrained optimization problem, for $j = 1, 2$:

$$\underset{\mathbf{M}_1, \mathbf{M}_2}{\text{minimize}} \quad \mathcal{L}(\mathbf{M}_1, \mathbf{M}_2) \quad \text{subject to} \quad \{\mathbf{M}_0\} = \{\mathbf{M}_1\} \cup \{\mathbf{M}_2\}, \mathbf{M}_j \in \mathcal{C}_j, \tag{3}$$

with $\mathcal{L}(\mathbf{M}_1, \mathbf{M}_2) = \left\| \left[ \mathcal{L}(\mathbf{M}_0), \sqrt{\frac{\#(\mathbf{M}_1)}{\#(\mathbf{M}_0)}} \mathcal{L}(\mathbf{M}_1), \sqrt{\frac{\#(\mathbf{M}_2)}{\#(\mathbf{M}_0)}} \mathcal{L}(\mathbf{M}_2) \right] \right\|_\infty$. As before, the minimization is subject to constraints, $\mathcal{C}_j$, $j = 1, 2$, which can be chosen for each time-lapse survey separately.

To produce time-lapse sampling masks, we employ simulated annealing as proposed by Zhang et al. [2022] but with the following differences: *(i)* randomly perturbed masks are drawn for each survey independently; *(ii)* the compound objectives and constraints of Equation 3 are used; *(iii)* to be relocated sample points are allowed to move more freely than during jitter sampling, which allows us to better explore candidate sampling masks. Figure 2 illustrates how the algorithm progresses when initialized with a replicated jittered subsampled (removing 80% of the sources) acquisition. From Figure 2a, we observe that the co-located source positions (denoted by the black dots) are gradually replaced by non-coincident source locations for the baseline (blue dots) and monitor surveys (red dots). Even though the objective of Equation 3 decreases non-monotonically (see Figure 2b), the reconstruction SNR increases for the baseline and monitor surveys for the selected points.

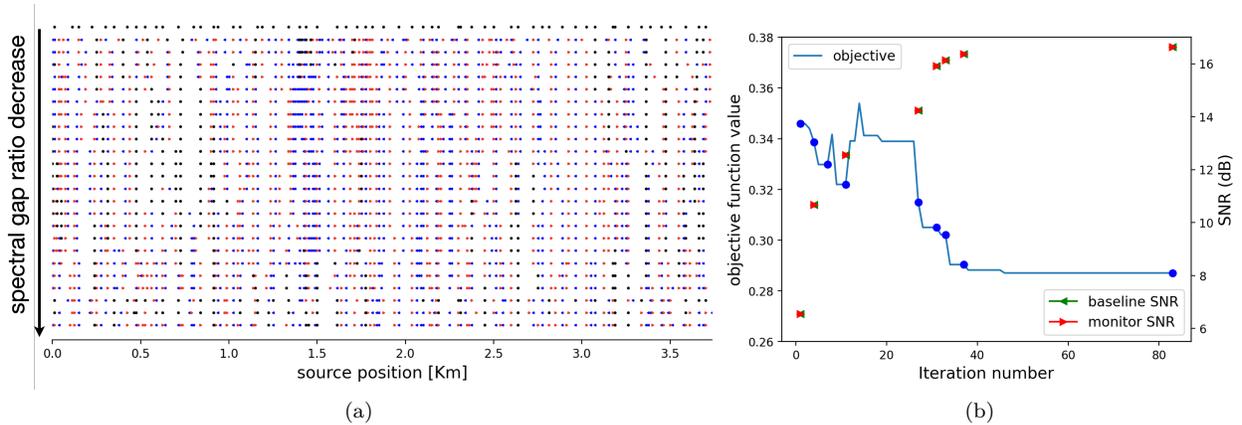

Figure 2: Automatic time-lapse sampling mask generation. (a) Starting from a jittered replicated sampling mask, the algorithm produces masks that have smaller SGRs but are no longer replicated. (b) Non-monotonically decaying objective and reconstruction SNR evaluated at points where the objective decreased by more than 0.003.

## Numerical validation

To confirm the benefits of optimized acquisition, we consider time-lapse data, which differs by a complex gas cloud [Wason et al., 2017, Jones et al., 2012]. Using finite-differences [Witte et al., 2019, Louboutin et al., 2022, 2019, Luporini et al., 2020], fully sampled (split spread) 2D baseline and monitor surveys are simulated each consisting of 300 sources/receivers sampled at 12.5 m. By using a single jittered subsampling mask 80% of the sources are removed, yielding an average source sampling rate of 62.5 m with 100% overlap. After running the optimization, the SGRs of the baseline/monitor surveys decreases from 0.346 to 0.268 and 0.262, respectively. The reduction in the overlap ratio (to 22%) leads to improvement in wavefield recovery via matrix completion [Kumar et al., 2015, 2017], which results in better SNRs for the baseline from 6.55 dB to 17.03 dB and for the monitor from 6.67 dB to 16.99 dB. For reasons explained by Oghenekohwo et al. [2017] and Wason et al. [2017], time-lapse difference plots are not included because the benefits of exact replication vanish when acquisition geometries undergo relatively small $(1-2\text{m})$ random shifts.

While these improvements are encouraging, the proposed optimization is approximate and the produced binary masks will be different for different starting masks. To investigate this effect, 30 overlapping jittered masks are generated by removing 75% of the sources. By reducing the overlap to $29\% \pm 8\%$, the optimized masks improve the SGRs as can be observed from the violin plots in Figure 4a. As before, the reductions in SGRs translate into improved SNRs as can be seen in Figure 4b. Compared to box plots, violin plots display the entire distribution including lines for the median (long dashes), first and third quartile (short dashes). We can make the following observations: *(i)* the SGRs for the baseline/monitor surveys decrease significantly; *(ii)* because of the larger number of sampled sources, the SGR for the common component is smaller and more narrowly distributed; *(iii)* the distribution of the SGRs of the baseline/monitor surveys is also narrow compared to the one of the initial jittered binary sampling masks; *(iv)* the SNRs for the recovered baseline/monitor surveys improve significantly.

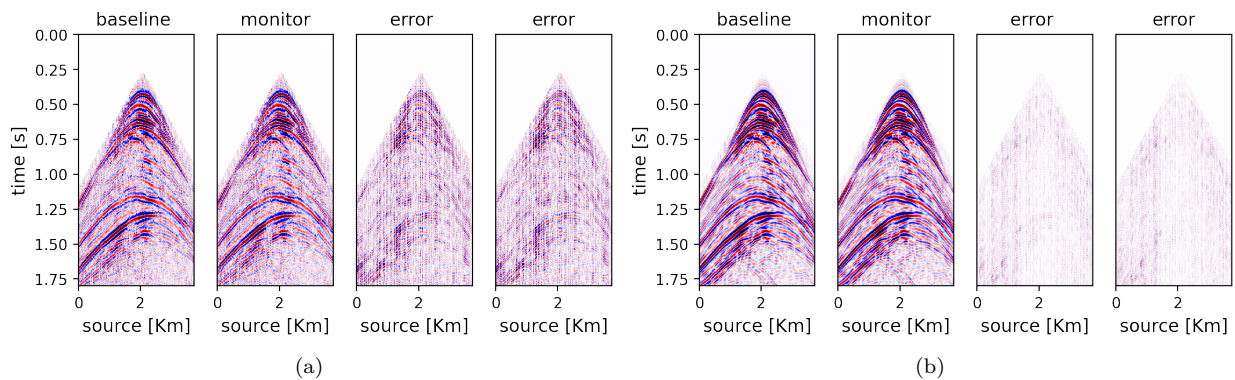

Figure 3: Time-lapse wavefield reconstruction in the time-domain. (a) wavefield reconstruction from 80% jittered subsampling for the baseline SNR = 6.55 dB, monitor SNR = 6.67 dB, and errors between the ground truth and the reconstructed wavefields. (b) the same but with optimized sampling masks, yielding improved recovery baseline/monitor surveys with SNR = 17.03, 16.99dB, respectively.

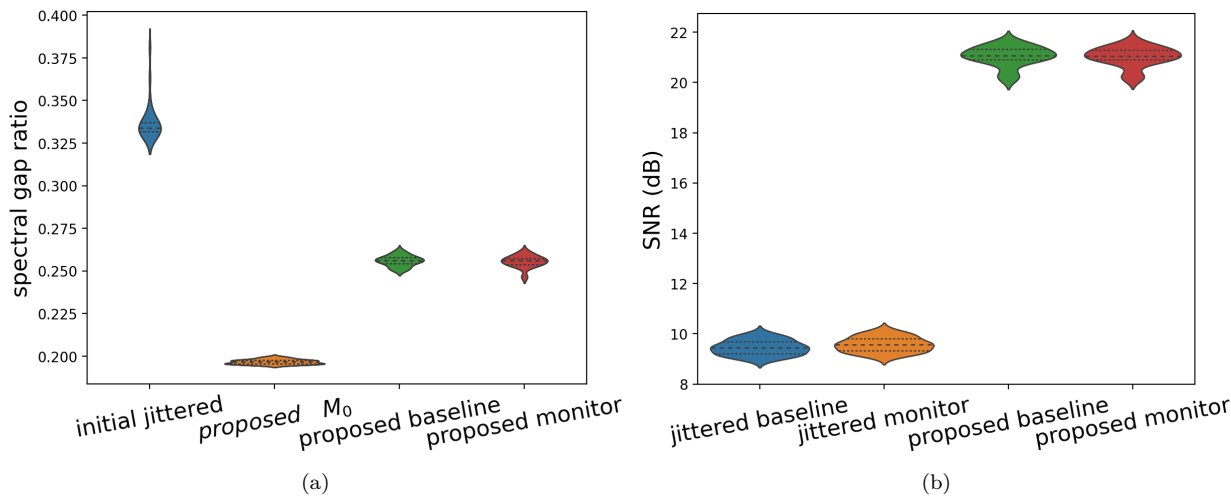

Figure 4: Violin plots for the SGRs (a) and recovery SNRs (b) for 30 independent experiments. These experiments show systematic reductions in SGR and significantly improves reconstruction SNRs for optimized surveys.

Even though the above results are encouraging and consistent with published reports that claim benefits of the JRM [Oghenekohwo et al., 2017, Wason et al., 2017, Yin et al., 2023], further scrutiny is in order. To this end, additional experiments were conducted to better understand robustness of the proposed methodology. Aside from predictable behavior for different starting masks (Figure 4), we also found that optimized SGRs are relatively insensitive to different runs of SA and to random perturbations in the optimized masks. The first observation implies that while SA may produce different masks, the SGRs remain very close, yielding wavefield reconstructions of near equal quality. The second observation indicates that postplot errors by single gridpoint shifts (12.5m) in the worst scenario offset the gains made by the optimization. However, on average improvements are mostly preserved although with higher variability.

The observed robustness of the presented method is consistent with reported behavior of the JRM. Even though we only considered the on-the-grid case, the argument can be made that improvements will carry over to the off-the-grid situation [Wason et al., 2017, Oghenekohwo and Herrmann, 2017, López et al., 2016]. However, to turn this claim into a more solid argument, we would have to extend the presented approach to the infinite-dimensional case, which is beyond the scope of this letter.

# Conclusions

Acquisition costs form a major impediment to time-lapse seismic. To reduce these costs while ensuring time-lapse repeatability, a novel acquisition optimization scheme was proposed that produces binary sampling masks that favor wavefield reconstruction with the joint recovery model. Optimized sampling masks were generated automatically by minimizing a new objective function consisting of spectral gap ratios for the baseline/monitor surveys and for the common component shared by the surveys. Aside from allowing for wave-simulation free, and therefore computationally feasible, optimized acquisition design, the proposed method also reaffirms the suggestion that deliberate relaxation of survey replication may lead to improved quality of jointly inverted surveys. This claim is solely based on connectivity arguments for the acquisition geometries associated with the baseline/monitor surveys and the common component. Because the spectral gap ratio is extremely cheap to evaluate, it lends itself very well to be extended to multiple monitoring surveys and to 3D. Off-the-grid acquisition geometries are also conducive to being improved by spectral gap ratios, but we will leave that extension to future work.